\documentclass[twoside,leqno,twocolumn]{article}
\usepackage{siamproceedings}

\usepackage[T1]{fontenc}
\usepackage{amsfonts}
\usepackage{graphicx}
\usepackage{epstopdf}
\usepackage{enumitem}
\usepackage{algorithmic}
\usepackage{seqsplit}        
\newcommand{\smiles}[1]{\seqsplit{\texttt{#1}}}

\ifpdf
  \DeclareGraphicsExtensions{.eps,.pdf,.png,.jpg}
\else
  \DeclareGraphicsExtensions{.eps}
\fi

\newsiamremark{remark}{Remark}
\newsiamremark{hypothesis}{Hypothesis}
\crefname{hypothesis}{Hypothesis}{Hypotheses}
\newsiamthm{claim}{Claim}

\makeatletter
\renewcommand{\ALG@name}{\sc Algorithm}
\makeatother

\usepackage{amsopn}

\usepackage{pifont}
\usepackage{tabularx,booktabs}
\usepackage{amssymb} 

\usepackage[colorinlistoftodos,prependcaption,textsize=tiny]{todonotes}

\begin{document}

\title{\Large Continuous Petri Nets for Fast Yield Computation: Polynomial-Time and MILP Approaches}

    \author{Addie Jordon\thanks{Universit\"at Bielefeld, Bielefeld, Germany (\email{addie.jordon@uni-bielefeld.de})}
    \and Juri Kol\v{c}\'ak\footnotemark[2]  \thanks{Universit\"at Bielefeld, Bielefeld, Germany (\email{juri.kolcak@uni-bielefeld.de})}
    \and Daniel Merkle\thanks{Universit\"at Bielefeld, Bielefeld, Germany (\email{daniel.merkle@uni-bielefeld.de})}}

\date{}

\maketitle



\begin{abstract}
Petri nets provide accurate analogues to chemical reaction networks, with places representing individual molecules (the resources of the system) and transitions representing chemical reactions which convert educt molecules into product molecules. Their natural affinity for modeling chemical reaction networks is, however, impeded by their computational complexity, which is at least \textsc{PSpace}-hard for most interesting questions, including reachability. Continuous Petri nets offer the same structure and discrete time as discrete Petri nets, but use continuous state-space, which allows them to answer the reachability question in polynomial time.
We exploit this property to introduce a polynomial time algorithm for computing the maximal yield of a molecule in a chemical system.
Additionally, we provide an alternative algorithm based on mixed-integer linear programming with worse theoretical complexity, but better runtime in practice, as demonstrated on both synthetic and chemical data.


\end{abstract}

\section{Introduction.}

The ability to determine if a particular compound can be produced from a set of starting molecules and, if it can be produced, identify the maximum yield, is of interest in chemistry and biosynthesis. Answering these questions in the context of metabolic networks is especially important as it allows one to identify which enzymes are involved in the synthesis of a given metabolite and which enzymes aid in increasing the yield. The answers to these questions are therefore highly relevant to synthetic biology.

Typical approaches to metabolic modeling include ordinary differential equations (ODEs) and flux balance analysis (FBA) \cite{orth2010flux}. ODEs have long been used in metabolomics for modeling metabolite concentrations over time \cite{goryanin1999mathematical, voit2017best}. ODE models require knowledge of a large number of precise kinetic parameters (e.g.~reaction rates); for large and complex systems such as metabolic networks, knowledge of such parameters is often incomplete or missing. FBA mitigates that requirement by assuming a steady state (wherein the concentration of molecules is consistently maintained) and establishing an objective function which typically maximizes biomass. As a result, the model is reduced to a linear program. However, FBA is unsuitable for analysis in a constrained setting where the starting resources are limited, such as when maximizing the yield of product molecule(s) from given source molecules. This is because FBA assumes that all molecules are always available in quantities sufficient to carry out any reaction.

Another model choice is Petri nets \cite{petri1962kommunikation}, which structurally provide a highly transparent analogue to chemical reaction networks, given their ability to model resource types (in our case, molecules) and the dynamics between them (in our case, chemical reactions). \emph{Discrete Petri nets}, often referred to simply as `Petri nets', originate from the thesis of Carl Adam Petri \cite{petri1962kommunikation}. They are a notable choice for modeling biological processes; for example, \cite{gupta2022validation} uses discrete Petri nets to model the biosynthesis of polyhydroxalkanoates (PHAs), and \cite{liu2021petri} presents a Petri nets-based framework for whole cell modeling.

Petri nets are directed bipartite graphs between two disjoint sets of vertices: \emph{places}, which represent the resource types, and \emph{transitions}, which define the ability to transform one resource type into another, in accordance to the \emph{arcs} connecting them to the places.



Classically, Petri nets are a discrete dynamical model. Each place can hold an amount of tokens representing the quantity of said resource type present in the system. The transitions can then \emph{fire} one at a time (chosen non-deterministically), given there are enough resources for them to consume, yielding discrete-time and discrete-state-space dynamics. Such dynamics are very faithful to the chemical reality, as molecules are fundamentally discrete entities. However, the complexity of discrete Petri net analysis (reachability has long been known to be at least \textsc{PSpace}-hard~\cite{esparza1998decidability} and has recently been proven to be Ackermann-complete for the related model of vector addition systems~\cite{czerwinski2022reachability}) is prohibitive to their application to many complex systems of interest, such as metabolic networks.

We thus propose continous Petri nets (CPNs) \cite{blondin2017logical} as a metabolic modeling approach, which opens the discrete state space dynamics of Petri nets to continuous state space. The CPN approach is relatively simple as it requires only three input variables: the metabolic network (molecules and reactions between them), the initial quantities of all molecules in the network, and the compound whose yield should be maximized. It has been shown that reachability can be solved polynomially using CPNs \cite{blondin2024separators} and a polynomial time algorithm for reachability has been introduced~\cite{fraca2015complexity}, which we implement as part of our solution.

The incidence matrix of the Petri net corresponds exactly to the stoichiometric matrix of the underlying chemical reaction network, which is the central structure used in FBA.
When using continuous state-space, FBA analysis thus becomes a special case of CPN analysis. By demanding that the steady state is preserved and that the net effect of all executed reactions is zero, one searches for \emph{transition invariants} (T-invariants) of the CPN.



By demanding a steady state, FBA assumes that every molecule is present in sufficient concentration; as a result, the order in which the reactions are executed becomes irrelevant as the educts of any reaction are present at any time. In contrast, the resource-limited setting of our CPN approach requires a check of \emph{causal soundness}, i.e.~that the educts of each reaction are present or can be produced in sufficient quantities before the reaction is executed. Causal soundness can also be verified in polynomial time and is indeed part of the reachability algorithm in \cite{fraca2015complexity}.

Such causal soundness is of special interest in metabolic networks. Since transitions in a CPN represent chemical reactions, a causally sound solution can be seen as an ordered list of reactions which are essential for optimal synthesis. From there, it is is possible to remove certain reactions and test how the yield changes.

Our approach consists of two separate solutions. The first is an algorithm \textsc{AtLeastReachable} which decides in polynomial time if at least $x \in \mathbb{R}_0^+$ token mass can be put on a single goal place from an initial marking. If possible, the algorithm returns the amount that each transition must fire. Although a witness is not returned, this solution is guaranteed to be causally sound. By wrapping \textsc{AtLeastReachable} in \textsc{BinarySearch}, one can determine the maximum possible yield.

\textsc{AtLeastReachable} achieves polynomial runtime by utilizing as many transitions as possible, producing large solutions which often contain spurious transitions (ie.~transitions which are not necessary to achieve the target yield).
In chemical networks, the `minimal' solution (the smallest set of transitions which \textbf{must} be fired in order to attain the goal mass) is often desirable, especially when the goal is to infer an underlying chemical mechanism. For that reason, we provide a secondary solution which trades polynomial time for minimal solution sets by using mixed-integer linear programming (MILP).

The paper is structured as follows: continuous Petri net background and terminology is covered in Sec.~\ref{sec:background}; algorithms for maximizing token mass (\textsc{AtLeastReachable} and \textsc{MILPMax}) are presented in Sec.~\ref{sec:methods} alongside proofs of correctness and polynomial running time for \textsc{AtLeastReachable}; modeling chemical reaction networks using continuous Petri nets is covered in Sec.~\ref{sec:results}, along with an example application and analysis of carbon efficiency in the pentose phosphate pathway, and comparative runtimes of \textsc{AtLeastReachable} with \textsc{Binary Search} and \textsc{MILPMax} in practice; and lastly a summary is provided in Sec.~\ref{sec:summary}. The open-source implementation associated with this paper is accessible at \url{https://github.com/a2390yu/cpns-a}.

\section{Background.} \label{sec:background}

The following formal definitions are based on notation and terminology given in \cite{haddad2024expressive}.

\subsection{Continuous Petri nets.}

Continuous Petri nets (CPNs) \cite{alla1998continuous} are structurally identical to discrete Petri nets; both are bipartite directed graphs on two disjoint sets of vertices (places and transitions). Places and transitions are connected by weighted arcs, where each weight defines a ratio between consumed and produced token mass.


\begin{definition}[Continuous Petri net (CPN), Petri net]

A CPN is a four-tuple $\mathcal{N} = (P, T, \texttt{In}, \texttt{Out})$ such that:

\begin{itemize}
    \item $P$ is a finite set of places;
    \item $T$ is a finite set of transitions, $T\cap P=\emptyset$;
    \item \texttt{In} and \texttt{Out} are the backward and forward incidence matrices respectively, which describe the incoming and outgoing arcs between place and transition tuples: $(P \times T) \rightarrow \mathbb{N}$.
\end{itemize}

\end{definition}

\begin{figure}[ht]
    \centering
    \includegraphics[width=\columnwidth]{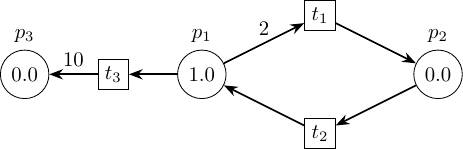}
    \caption{An example CPN. The places are depicted as circles and transitions as squares, as per convention. The arc weights equal to $1$ are omitted. The CPN is marked with an initial marking depicted by the token mass in each individual place.}\label{fig:cpnex}
\end{figure}

Fig.~\ref{fig:cpnex} shows an example CPN with three places $P=\{p_1,p_2,p_3\}$ and three transitions $T=\{t_1,t_2,t_3\}$. Each weighted arc connects places to a transition -- or a transition to places -- and visualizes the values of the backward and forward incidence matrices, $\texttt{In}$ and $\texttt{Out}$ respectively, given below.

\vskip 12pt
\parbox{0.4\columnwidth}{%
    \begin{tabular}{ r | c c c }
        \texttt{In} & $p_1$ & $p_2$ & $p_3$ \\
        \hline
        $t_1$ & 2 & 0 & 0 \\
        $t_2$ & 0 & 1 & 0 \\
        $t_3$ & 1 & 0 & 0 \\
    \end{tabular}
}%
\hskip 2em
\parbox{0.4\columnwidth}{%
    \centering%
    \begin{tabular}{ r | c c c }
        \texttt{Out} & $p_1$ & $p_2$ & $p_3$ \\
        \hline
        $t_1$ & 0 & 1 & 0 \\
        $t_2$ & 1 & 0 & 0\\
        $t_3$ & 0 & 0 & 10 \\
    \end{tabular}
}%
\vskip 12pt

The \textit{incidence matrix} of a CPN is $C = \texttt{Out} - \texttt{In}$.
For any place $p\in P$, we use ${}^\bullet p=\{t \in T : \texttt{Out}(p,t) > 0\}$ and $p^\bullet = \{t \in T : \texttt{In}(p,t) > 0\}$ to represent the transitions producing and consuming token mass from $p$, respectively.
Similarly, for any $t\in T$, ${}^\bullet t = \{p \in P : \texttt{In}(p,t) > 0\}$ and $t^\bullet = \{p \in P : \texttt{Out}(p,t) > 0\}$ represent the input and output places of the transition $t$.
Given $T'\subseteq T$, $\mathcal{N}_{T'}$ is the net $\mathcal{N}$ restricted to transitions in $T'$, with place set ${}^\bullet T'{}^\bullet$ and incidence matrices $\texttt{In}_{{}^\bullet T'{}^\bullet \times T'}$ and $\texttt{Out}_{{}^\bullet T'{}^\bullet \times T'}$.
The reverse of a CPN $\mathcal{N}$ is denoted $\mathcal{N}^{-1} = (P, T, \texttt{Out}, \texttt{In})$.


\begin{definition}[Marking, marked CPN, initial marking]
    A marking of a CPN $\mathcal{N}=(P,T,\texttt{In},\texttt{Out})$ is a vector $\mathbf{m} \in {\mathbb{R}^+_0}^P$. The CPN coupled with a marking $\mathbf{m}_0$, $(\mathcal{N}, \mathbf{m}_0)$, is a \textit{marked CPN} and the marking $\mathbf{m}_0$ is the initial marking. We use $\mathbf{m}(p)$ to retrieve the amount of token mass on place $p$ in marking $\mathbf{m}$.
\end{definition}

A marking assigns a non-negative real amount of token mass to each place, thus capturing the state of the system.
The CPN in Fig.~\ref{fig:cpnex} is marked with the initial marking $\mathbf{m}_0=(p_1\colon 1,p_2\colon 0,p_3\colon 0)$.

\begin{definition}[Enabling degree]
     Given a CPN $\mathcal{N}$ and a marking $\mathbf{m}$, the enabling degree of transition $t\in T$ in marking $\mathbf{m}$ is defined as $\texttt{enab}(t, \mathbf{m}) = \underset{p \in ^\bullet t}{\texttt{min}}(\frac{\mathbf{m}(p)}{\texttt{In}(p,t)})$ when $^\bullet t \neq \emptyset$ and $\infty$ otherwise.
     We say a transition $t\in T$ is enabled in $\mathbf{m}$ if $\texttt{enab}(t,\mathbf{m})>0$.
\end{definition}

The enabling degree specifies the maximum amount of token mass that can be moved via a transition (in the given marking) and is bounded by the input place with the least available weighted token mass. In the example in Fig.~\ref{fig:cpnex}, both transitions $t_1$ and $t_3$ are enabled with degrees $0.5$ and $1$, respectively. However, $t_2$ is not enabled, since $p_2\in{}^\bullet t_2$ and $\mathbf{m}_0(p_2)=0$.


\begin{definition}[Firing]
    An enabled transition $t\in T$ can \textit{fire} by any amount $\alpha \in [0, \texttt{enab}(t,\mathbf{m})]\cap\mathbb{R}$, resulting in a new marking $\mathbf{m}' = (\mathbf{m}(p)+\alpha C(p,t), \forall p \in P)$, written as $\mathbf{m} \overset{\alpha t}{\rightarrow}_\mathcal{N}\mathbf{m}'$. When $\alpha=1$, it is sometimes omitted in notation, yielding $\mathbf{m} \overset{t}{\rightarrow}_\mathcal{N}\mathbf{m}'$
\end{definition}

In Fig.~\ref{fig:cpnex}, $t_3$ can fire with any $\alpha\in[0,1]$, resulting in a marking $\mathbf{m}=(p_1\colon 1-\alpha, p_2\colon0,p_3\colon 10\alpha)$, with a ten-fold token mass on $p_3$.
In chemistry, this could correspond to a large molecule ($p_1$) undergoing a fragmentation process ($t_3$) to result in a larger quantity of smaller molecules ($p_3$).
Similarly, $t_1$ can also be fired with any $\alpha\in[0,0.5]$.
Crucially, both $t_3$ and $t_1$ can be fired in sequence, provided $t_3$ is fired with $\alpha<1$ and $t_1$ with $\alpha<0.5$.

\begin{definition}[Firing sequence]
    Let $\mathcal{Z}=\mathbb{R}^+_0 \times T$ denote the set of \textit{firing steps} whose members are written as $\alpha t$. Let $\sigma = (\alpha_i t_i)_{i\leq n}$ be a finite sequence over $\mathcal{Z}$ of length $n\in\mathbb{N}$. Then $\sigma$ is a \textit{finite firing sequence} if there exists a finite sequence of markings $(\mathbf{m}_i)_{i\leq n+1}$ such that for all $i \leq n$, $\mathbf{m}_i \overset{\alpha_i t_i}{\rightarrow}\mathbf{m}_{i+1}$. We write such a firing sequence as $\mathbf{m}_0 \overset{\sigma}{\rightarrow}_\mathcal{N}\mathbf{m}_{n+1}$. From this point onward, let the term \textit{firing sequence} refer to \textit{finite firing sequence} unless otherwise specified.
\end{definition}

A firing sequence states that it is possible to reach a marking $\mathbf{m}_{f}$ from $\mathbf{m}_0$ through a sequence of at most $n \in \mathbb{N}$ firings.
We write $\mathbf{m}_0 \overset{\sigma}{\rightarrow}_\mathcal{N}$ in case the final marking is not important.

\begin{definition} [Infinite firing sequence]
     Let $\sigma = (\alpha_i t_i)_{i\in\mathbb{N}}$ be an infinite sequence over $\mathcal{Z}$. Then $\sigma$ is called an \textit{infinite firing sequence} if there exists an infinite family of markings $(\mathbf{m}_i)_{i\leq\omega}$ such that $\mathbf{m}_i \overset{\alpha_i t_i}{\longrightarrow} \mathbf{m}_{i+1}$ $\forall i \in \mathbb{N}$ and $\lim_{i\rightarrow\infty} \mathbf{m}_i = \mathbf{m}_\omega$. The infinite firing sequence is then written as $\mathbf{m}_0 \overset{\sigma}{\rightarrow}_\mathcal{N} \mathbf{m}_\omega$.
\end{definition}

Infinite firing sequences express the limit behavior of CPNs.
Firing $t_1$ and subsequently $t_2$ with the same $\alpha$ in the CPN from Fig.~\ref{fig:cpnex} reduces the token mass on $p_1$ by $\frac{\alpha}{2}$.
After any finite amount of firings of the $t_1$ and $t_2$ loop, some token mass is guaranteed to remain on either $p_1$ or $p_2$,
but an infinite firing sequence resulting in the empty marking $\mathbf{m}=\mathbf{0}$ exists, e.g.~$(2^{-n}t_12^{-n}t_2)_{n\in\mathbb{N}}$, which resembles a damped oscillation.

\begin{definition}[Reachability, reachable]
    Consider the marked CPN $(\mathcal{N}, \mathbf{m}_0)$. The \textit{reachability set} is defined as $\textbf{RS}(\mathcal{N}, \mathbf{m}_0) = \{\mathbf{m} : \exists \sigma \in \mathcal{Z}^*, \mathbf{m}_0 \overset{\sigma}{\rightarrow}\mathbf{m}\}$. A marking $\mathbf{m}_r$ is said to be \textit{reachable} from $\mathbf{m}_0$ iff.~$\mathbf{m}_r \in \textbf{RS}(\mathcal{N}, \mathbf{m}_0)$.
\end{definition}

\begin{definition}[Limit-reachability, limit-reachable]
    Consider the marked CPN $(\mathcal{N}, \mathbf{m}_0)$. The \textit{limit-reachability set} is defined as $\texttt{lim-}\textbf{RS}(\mathcal{N}, \mathbf{m}_0) = \{\mathbf{m} : \exists \sigma \in \mathcal{Z}^\omega, \mathbf{m}_0 \overset{\sigma}{\rightarrow}\mathbf{m}\}$. A marking $\mathbf{m}_r$ is said to be \textit{limit-reachable} from $\mathbf{m}_0$ iff.~$\mathbf{m}_r \in \texttt{lim-}\textbf{RS}(\mathcal{N}, \mathbf{m}_0)$.
\end{definition}

Note that the associated firing sequence of a marking in the limit-reachability set must be infinite. However, as transitions are allowed to fire with $\alpha=0$, the reachability set is a subset of the limit-reachability set.
The subset relation $\mathbf{RS}(\mathcal{N},\mathbf{m}_0)\subseteq\texttt{lim-}\mathbf{RS}(\mathcal{N},\mathbf{m}_0)$ is generally strict, as illustrated by the example in Fig.~\ref{fig:cpnex}, $\mathbf{0}\in\texttt{lim-}\mathbf{RS}(\mathcal{N},\mathbf{m}_0)\setminus\mathbf{RS}(\mathcal{N},\mathbf{m}_0)$.

\begin{definition}[Support]
For a vector $v$, let $v^+$ denote the \textit{support} of $v$, with $v^+=\{ x : x \in v; x > 0\}$.
\end{definition}

\begin{definition}[Parikh image]
Given an (infinite) firing sequence $\sigma$, the \textit{Parikh image} of $\sigma$ is $\overset{\rightarrow}{\sigma}=(t\in T : \sum_{t_i = t} \alpha_i)$.
\end{definition}

For each transition $t \in T$, the Parikh image sums the firing intensities of all instances of $t$ along the (infinite) firing sequence.
The type of a Parikh image is thus $(\mathbb{R}_0^+\cup\{\infty\})^T$.

\begin{definition}[Firing set] \label{firing-set}
    The \textit{firing set} of a marked CPN $(\mathcal{N}, \mathbf{m}_0)$ is $\textbf{FS}(\mathcal{N},\mathbf{m}_0) = \{\overset{\rightarrow}{\sigma}^+ : \exists \sigma\in \mathcal{Z}^*, \mathbf{m}_0 \overset{\sigma}{\rightarrow}_\mathcal{N}\}$.
\end{definition}

A set of transitions $T'\subseteq T$ belongs to the firing set $\textbf{FS}(\mathcal{N},\mathbf{m}_0)$ iff there exists a firing sequence $\sigma$ which uses exactly the transitions in $T'$; its size is therefore on the order of $O(2^{|T|})$. The firing amounts are irrelevant as long as they are non-zero, as only the support of the Parikh vector is considered. In this way, each element of the firing set can be viewed as a Boolean vector $\mathbf{b}\in\mathbb{B}^T$ such that $\mathbf{b}_t =1$ if $t\in T'$ and $0$ otherwise for each $t\in T$. We make use of this Boolean variable view of transitions in the MILP formulation.

The notion of firing sets is paramount for determining causal soundness.
Indeed, by Theorems 19 and 20 of~\cite{fraca2015complexity}, $\mathbf{m}\in\mathbf{RS}(\mathcal{N},\mathbf{m}_0)$ (respectively $\mathbf{m}\in\texttt{lim-}\mathbf{RS}(\mathcal{N},\mathbf{m}_0)$) is equivalent to existence of a vector $\mathbf{v}\in{\mathbb{R}_0^+}^T$ satisfying the following conditions:
\begin{enumerate}
    \item $\mathbf{m}=\mathbf{m}_0+C\mathbf{v}$;
    \item $\mathbf{v}^+\subseteq \mathbf{FS}(\mathcal{N},\mathbf{m}_0)$;
    \item $\mathbf{v}^+\subseteq \mathbf{FS}(\mathcal{N}^{-1},\mathbf{m})$ (only for $\mathbf{m}\in\mathbf{RS}(\mathcal{N},\mathbf{m}_0)$);
\end{enumerate}

The vector $\mathbf{v}$ gives the collective amounts each transition should fire to attain $\mathbf{m}$, and can be computed using LP with the matrix equation $C\mathbf{v}=\mathbf{m}-\mathbf{m}_0$.
The check against the firing set is then necessary to determine whether $\mathbf{v}$ represents a causally sound solution.

\begin{figure}[ht]
    \centering
    \includegraphics[width=0.6\columnwidth]{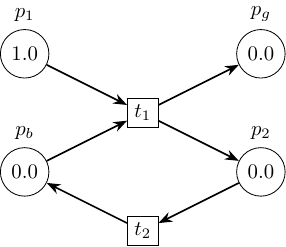}
    \caption{A marked CPN whose matrix equation permits putting token mass $1$ on the place $p_g$, but whose firing set only contains the empty set.}\label{fig:cpn-unreal}
\end{figure}

Consider the CPN in Fig.~\ref{fig:cpn-unreal}, with the initial marking $\mathbf{m}_0=(p_1:1,p_2:0,p_b:0,p_g:0)$. To maximize token mass on $p_{g}$, a valid solution to the LP is $\mathbf{v} = (t_1: 1.0, t_2:1.0)$. To explain at a high level, token mass is `borrowed' from $p_b$ and used to fire $t_1$, resulting in token mass on $p_g$ and $p_2$. The token mass on $p_2$ is then restored to $p_g$ via $t_2$, resulting in a net change of zero token mass for $p_b$. 

However, such a solution is not causally sound, since $p_b$ has no initial token mass; in other words, the set $\{t_1, t_2\}$ is not a member of the firing set.
We refer to such LP solutions which are not causally sounds as \emph{unrealizable}, i.e.~solutions for which no (infinite) firing sequence exists which would fire each transition to the full amount specified in the solution vector.

\section{Methodology.} \label{sec:methods}

\subsection{Deciding achievable token mass.}

The algorithms for deciding membership in the firing set (\textsc{Fireable}, Alg.~\ref{alg:fireable}) and deciding exact reachability (\textsc{Reachable}, Alg.~\ref{alg:reachable}) are adapted from \cite{fraca2015complexity} and serve as the basis for our implementation.

\begin{algorithm}[ht]
    \small
    \caption{Decision algorithm for membership of $FS(\mathcal{N}, \mathbf{m}_0)$ \cite{fraca2015complexity}}\label{alg:fireable}
    \hspace*{\algorithmicindent} \texttt{Fireable}($\langle\mathcal{N}, \mathbf{m}_0\rangle$, $T'$):\\
    \hspace*{\algorithmicindent} \textbf{Input}: a CPN system $\langle\mathcal{N}, \mathbf{m}_0\rangle$, a subset of transitions~$T'$ \\
    \hspace*{\algorithmicindent} \textbf{Output}: the membership status of $T'$ w.r.t.~$FS(\mathcal{N}, \mathbf{m}_0)$ \\
    \hspace*{\algorithmicindent} \textbf{Output}: in the negative case, the maximal firing set included in $T'$
    \begin{algorithmic}[1]
        \STATE{$T'' \gets \emptyset; P' \gets \mathbf{m}_0^+$}
        \WHILE{$T'' \neq T'$}
            \STATE{$new \gets $ \textbf{false}}
            \FOR {$t \in T' \setminus T''$}
                \IF{$^\bullet t \subseteq P'$}
                    \STATE{ $T'' \gets T'' \cup \{t\}$}
                    \STATE{$P' \gets P' \cup t^\bullet$}
                    \STATE{$new \gets$ \textbf{true}}
                \ENDIF
            \ENDFOR
            \IF{\textbf{not} $new$}
                \RETURN{(\textbf{false}, $T''$)}
            \ENDIF
        \ENDWHILE
        \RETURN{\textbf{true}}
    \end{algorithmic}
\end{algorithm}

\begin{algorithm}[ht]
    \small
    \caption{Decision algorithm for reachability \cite{fraca2015complexity}}\label{alg:reachable}
    \hspace*{\algorithmicindent} \texttt{Reachable}($\langle\mathcal{N}, \mathbf{m}_0\rangle$, $\mathbf{m}$):\\
    \hspace*{\algorithmicindent} \textbf{Input}: a CPN system $\langle\mathcal{N}, \mathbf{m}_0\rangle$, a marking $\mathbf{m}$ \\
    \hspace*{\algorithmicindent} \textbf{Output}: the reachability status of $\mathbf{m}$ \\
    \hspace*{\algorithmicindent} \textbf{Output}: the Parikh image of a witness in the positive case
    \begin{algorithmic}[1]
        \IF{$\mathbf{m}=\mathbf{m}_0$}
            \RETURN{(\textbf{true, 0})}
        \ENDIF
        \STATE{$T' \gets T$}
        \WHILE{$T' \neq \emptyset$}
            \STATE{$nbsol \gets 0$}
            \STATE{$\textbf{sol} \gets 0$}
            \FOR{$t \in T'$}
                \STATE{\textbf{solve} $\exists?\mathbf{v}$ $\mathbf{v} \geq \textbf{0} \land \mathbf{v}[t] > 0 \land C_{P\times T'}\mathbf{v} = \mathbf{m} - \mathbf{m}_0$} \label{rea:line:lp}
                \IF{$\exists \mathbf{v}$}
                    \STATE{$nbsol \gets nbsol + 1$}
                    \STATE{$\textbf{sol} \gets \textbf{sol}+\mathbf{v}$}
                \ENDIF
            \ENDFOR
            \IF{$nbsol = 0$}
                \RETURN{(\textbf{false}, $T''$)}
            \ELSE{
                \STATE{$\textbf{sol} \gets \frac{1}{nbsol} \textbf{sol}$}
                }
            \ENDIF
            \STATE{$T' \gets \textbf{sol}^+$} \label{rea:line:sol}
            \STATE{$T' \gets T' \cap \texttt{maxFS} (\mathcal{N}_{T'}, \mathbf{m}_0[^\bullet T'^\bullet])$} \label{rea:line:fs}
            \STATE{$T' \gets T' \cap \texttt{maxFS} (\mathcal{N}_{T'}^{-1}, \mathbf{m}_0[^\bullet T'^\bullet])$ \texttt{/* deleted for lim-reachability */}} \label{rea:line:fs_fin}
            \IF{$T'= \textbf{sol}^+$}
                \RETURN{(\textbf{true, sol})}
            \ENDIF
        \ENDWHILE
        \RETURN{\textbf{false}}
    \end{algorithmic}
\end{algorithm}

For a given subset of transitions $T' \subseteq T$ of a marked CPN $(\mathcal{N}, \mathbf{m}_0)$, \textsc{Fireable} decides whether $T' \in \mathbf{FS}(\mathcal{N}, \mathbf{m_0})$. In the event that $T'$ is not in the firing set, the largest subset $T''\subset T'$ in the firing set is returned alongside the boolean \textit{false} indicator.

\textsc{Reachable} decides whether a given marking is reachable (respectively, limit-reachable) in the input CPN.
The algorithm keeps track of a set of transitions $T'$, initially the whole $T$, which represents the support of a potential firing sequence to reach the target marking $\mathbf{m}$.
$T'$ becomes smaller in size in one of two ways: firstly, on line~\ref{rea:line:sol}, due to the aggregate linear program (line~\ref{rea:line:lp}) solution which ensures a linear combination of transitions $\mathbf{sol}$ exists which satisfies $\mathbf{m}_0+C\mathbf{sol}=\mathbf{m}$, thus over-approximating reachability; and secondly, by computing the intersection with the maximum firing set of the net $\mathcal{N}$ restricted to the current set $T'$ (line~\ref{rea:line:fs}). Such maximum firing sets can be computed using \textsc{Fireable}.

In the cases that either no solution to the linear program is found or the maximum firing set of $\mathcal{N}_{T'}$ becomes empty, the algorithm concludes that the marking is not reachable. On the other hand, if the linear program solution and the maximum firing set agree on a set of transitions $T'$, this set is outputted as the support of a witness of the reachability of $\mathbf{m}$. If only finite reachability is of interest, an extra check is enforced against the maximum firing set of the inverse net, line~\ref{rea:line:fs_fin}, as per Theorem 19 of~\cite{fraca2015complexity}.


\textsc{Reachable} decides reachability of the precise marking $\mathbf{m}$; that is, it answers the question `Is it possible to reach exactly the marking $\mathbf{m}$ from $\mathbf{m}_0$?'. However, our goal is to maximize token mass on a single `goal' place, and therefore the token mass on non-goal places is free to take on any non-negative real value. Thus, the construction of the linear program is adjusted in order to change the question from \textit{exact} to \textit{at least}; that is, instead of asking if an \textit{exact} marking $\mathbf{m}$ can be reached, we ask if \textit{at least} token mass $x \in \mathbb{R}_0^+$ on a (singular) goal place $p$ can be reached. More specifically, the linear program was changed from strict equality to an inequality: $\textbf{solve: } \exists?\mathbf{v}, \mathbf{v}\geq 0 \land \mathbf{v}[t] > 0 \land C_{P\times T'} \mathbf{v} \geq \mathbf{m}-\mathbf{m}_0$. Let us refer to this modified version of \textsc{Reachable} as \textsc{AtLeastReachable} (Alg.~\ref{alg:alreachable}).

\begin{algorithm}[ht]
    \small
    \caption{Decision algorithm for `at-least' reachability}\label{alg:alreachable}
    \hspace*{\algorithmicindent} \texttt{AtLeastReachable}($\langle\mathcal{N}, \mathbf{m}_0\rangle$, $\mathbf{m}$):\\
    \hspace*{\algorithmicindent} \textbf{Input}: a CPN system $\langle\mathcal{N}, \mathbf{m}_0\rangle$, a marking $\mathbf{m}$ whose support $\mathbf{m}^+$ is comprised only of goal place(s) \\
    \hspace*{\algorithmicindent} \textbf{Output}: the reachability status of $\mathbf{m}$ \\
    \hspace*{\algorithmicindent} \textbf{Output}: the Parikh image of a witness in the positive case
    \begin{algorithmic}[1]
        \IF{$\mathbf{m}\leq\mathbf{m}_0$}
            \RETURN{(\textbf{true, 0})}
        \ENDIF
        \STATE{$T' \gets T$}
        \WHILE {$T' \neq \emptyset$} \label{alr:line:while-loop}
            \STATE{$nbsol \gets 0$}
            \STATE{$\textbf{sol} \gets \textbf{0}$}
            \FOR{$t \in T'$} \label{alr:line:for-loop}
                \STATE{$\textbf{solve: } \exists?\mathbf{v}, \mathbf{v}\geq \textbf{0} \land \mathbf{v}[t] > 0 \land C_{P\times T'}\mathbf{v} \geq \mathbf{m}-\mathbf{m}_0$}
                \IF{$\exists \mathbf{v}$}
                    \STATE{$nbsol \gets nbsol + 1$}
                    \STATE{$\textbf{sol} \gets \textbf{sol}+\mathbf{v}$}
                \ENDIF
            \ENDFOR
            \IF{$nbsol = 0$}
                \RETURN{(\textbf{false}, $T''$)} \label{alr:line:false1}
            \ELSE
                \STATE{$\textbf{sol} \gets \frac{1}{nbsol} \textbf{sol}$}
            \ENDIF
            \STATE{$T' \gets \textbf{sol}^+$} \label{alr:line:sol+}
            \STATE{$T' \gets T' \cap \texttt{maxFS} (\mathcal{N}_{T'}, \mathbf{m}_0[^\bullet T'^\bullet])$} \label{alr:line:fs}
            \STATE{$T' \gets T' \cap \texttt{maxFS} (\mathcal{N}_{T'}^{-1}, \mathbf{m}_0[^\bullet T'^\bullet])$ \texttt{/* deleted for lim-reachability */}} \label{alr:line:fslim}
            \IF{$T'= \textbf{sol}^+$}
                \RETURN{(\textbf{true, sol})} \label{alr:line:return-true}
            \ENDIF
        \ENDWHILE
        \RETURN{\textbf{false}} \label{alr:line:false2}
    \end{algorithmic}
\end{algorithm}

It is important to note that the repeated construction and execution of a linear program for each $t \in T'$ (which, since $T$ is initialized to $T'$, is the same as $t \in T$) results in an aggregated solution variable. A side effect of relaxing the strict equality constraint in the original LP into a $\geq$ inequality constraint is a significant increase in spurious transitions added to the solution variable. Thus, solution vectors that result from \textsc{AtLeastReachable} are often large and contain spurious transitions, which do not contribute to the moving of token mass to the goal place. In the solution spurious transitions are often not only unneeded but unwanted as they can obscure the transitions which \textit{do} contribute to the goal place.

Note that \textsc{AtLeastReachable} must be used with another algorithm in order to determine the maximum amount of token mass placable on the goal place. We elect to use \textsc{BinarySearch} \cite{lin2019binary}, which repeatedly bisects an ordered search space in half until it converges upon an answer. The resulting complexity is a logarithmic number of calls (proportional to the desired level of precision) to the polynomial time algorithm.

The following proofs of correctness and runtime are largely adaptations of the equivalent proofs for \textsc{Reachable} given in~\cite{fraca2015complexity}.

\begin{proposition}
    \textsc{AtLeastReachable} returns true iff there exists a marking $\mathbf{m}'$, such that $\mathbf{m}'\geq\mathbf{m}$ and $\mathbf{m}'\in \mathbf{RS}(\mathcal{N}, \mathbf{m}_0)$. \textsc{AtLeastReachable} without line \ref{alr:line:fslim} returns true iff there exists a marking $\mathbf{m}'$, such that $\mathbf{m}'\geq\mathbf{m}$ and $\mathbf{m}'\in \texttt{lim-}\mathbf{RS}(\mathcal{N}, \mathbf{m}_0)$. 
\end{proposition}

\begin{proof}
    \textbf{Soundness.} We consider only the non-trivial case when $\mathbf{m}\not\leq \mathbf{m}_0$. Assume that \textsc{AtLeastReachable} returns with \textbf{true} at line \ref{alr:line:return-true}.
    Let $\mathbf{m}'=\mathbf{m}_0+C\mathbf{sol}$ where $\mathbf{sol}$ is the aggregate solution outputted by the algorithm. We have $C\mathbf{sol}\geq\mathbf{m}-\mathbf{m}_0$ since the inequality holds for each individual solution. Thus, $\mathbf{m}'\geq\mathbf{m}_0+\mathbf{m}-\mathbf{m}_0=\mathbf{m}$.
    
    Since $T'$ is assigned $\textbf{sol}^+$ on line \ref{alr:line:sol+}, this implies that lines \ref{alr:line:fs} and \ref{alr:line:fslim} do not change the value of $T'$, which in turn implies that $\textbf{sol}^+ \in FS(\mathcal{N}, \mathbf{m}_0)$ and $\textbf{sol}^+ \in FS(\mathcal{N}^{-1}, \mathbf{m})$. Thus, $\mathbf{m}'\in \mathbf{RS}(\mathcal{N},\mathbf{m}_0)$ as per Theorem 19 in \cite{fraca2015complexity}. Respectively, $\mathbf{m}'\in\texttt{lim-}\mathbf{RS}(\mathcal{N},\mathbf{m}_0)$ when line \ref{alr:line:fslim} is omitted, as per Theorem 20 in \cite{fraca2015complexity}.
\end{proof}

\begin{proof}
    \textbf{Completeness.}
    Assume \textsc{AtLeastReachable} returns \textbf{false}.
    We assert \textsc{AtLeastReachable} fulfills the following invariant at any time: for any $\mathbf{m}'\geq\mathbf{m}$ and any $\mathbf{m}_0\overset{\sigma}{\rightarrow}_{\mathcal{N}}\mathbf{m}'$, $\overset{\rightarrow}{\sigma}^+\subseteq T'$.
    The invariant holds initially since $T'=T$.
    By construction, for any $t\in T'$, $t\in \mathbf{sol}^+$ iff there exist $\mathbf{m}'\geq\mathbf{m}$ and $\mathbf{v}\in{\mathbb{R}_0^+}^T$ with $\mathbf{v}_t>0$ and $\mathbf{m}'=C\mathbf{v}+\mathbf{m}_0$. Thus the assignment at line \ref{alr:line:sol+} preserves the invariant as for any $\mathbf{m}'\geq \mathbf{m}$ and $\sigma$, $\overset{\rightarrow}{\sigma}^+\subseteq\mathbf{sol}^+$ as per Theorem 19 (respectively Theorem 20 in the limit-reachability case) of~\cite{fraca2015complexity}.

    Similarly, by Theorem 19 (respectively Theorem 20) of ~\cite{fraca2015complexity}, for any $\mathbf{m}'$ and $\sigma$, $\overset{\rightarrow}{\sigma}^+\subseteq \texttt{maxFS}(\mathcal{N}_{T'},\mathbf{m}_0[{}^\bullet T'{}^\bullet])$.
    The assignment on line~\ref{alr:line:fs} thus preserved the invariant.
    In case of finite reachability, Theorem 19 of~\cite{fraca2015complexity} extends also to line~\ref{alr:line:fslim}, for any $\mathbf{m}'$ and $\sigma$, $\overset{\rightarrow}{\sigma}^+\subseteq \texttt{maxFS}(\mathcal{N}_{T'}^{-1},\mathbf{m}_0[{}^\bullet T'{}^\bullet])$.

    If the algorithm returns \textbf{false} on line~\ref{alr:line:false1}, then by the invariant, the first condition of Theorem 19 (respectively Theorem 20) of~\cite{fraca2015complexity} cannot be satisfied for any $\mathbf{m}'\geq\mathbf{m}$.
    Finally, if the algorithm returns \textbf{false} on line~\ref{alr:line:false2}, then $T'=\emptyset$ and by $\mathbf{m}\not\leq\mathbf{m}_0$ and the invariant, for any $\mathbf{m}'\geq\mathbf{m}$, there exists no $\mathbf{m}_0\overset{\sigma}{\rightarrow}_{\mathcal{N}}\mathbf{m}'$ and thus $\mathbf{m}'\notin \mathbf{RS}(\mathcal{N},\mathbf{m}_0)$, respectively $\mathbf{m}'\notin \texttt{lim-}\mathbf{RS}(\mathcal{N},\mathbf{m}_0)$.

\end{proof}

\begin{proposition}
\textsc{AtLeastReachable} runs in polynomial time.
\end{proposition}

\begin{proof}
    The outer while-loop has at most $|T|$ iterations. $T'$ can only be modified on lines \ref{alr:line:sol+}--\ref{alr:line:fslim} and can only decrease in size through intersections. Additionally, $T'\subseteq T$, since $\mathbf{sol}$ is computed for the net restricted to $T'$. Furthermore, if $T'$ remains unchanged after lines \ref{alr:line:sol+}--\ref{alr:line:fslim}, the algorithm will terminate on line \ref{alr:line:return-true}.

    The inner for-loop is also bounded by $|T|$ since it runs once per each member of $T'$.

    Solving a linear program can be done in polynomial time \cite{papadimitriou1998combinatorial} as can computing the maximum firing set of a marked CPN \cite{fraca2015complexity}.
\end{proof}

\subsection{Mixed integer linear programming maximization.}

Although \textsc{AtLeastReachable} with \textsc{BinarySearch} runs in logarithmic iterations of polynomial time, the solution vectors it creates can be undesirable as they are large and often include spurious transitions. By using mixed-integer linear programming (MILP), we are able to obtain smaller solutions, that allow for an easier mechanistic interpretation. However, since MILP is NP-complete \cite{Karp1972}, the algorithm no longer runs in logarithmic iterations of polynomial time.

We construct a MILP with three main goals: (1) to maximize token mass on the desired place, (2) to retain the guarantee that solutions are members of the firing set, and (3) to prioritize smaller-size solutions, if multiple solutions share the same objective value. 

To directly maximize a goal place $p_g$, the objective function becomes $obj = {}^\bullet p_g - p_g ^\bullet$. Since finding the solution which fires the fewest number of transitions is desired, we use boolean variables to express the size of the solution; $\mathbf{b} = (b_t,  \forall t \in T)$ are used to represent which transitions are fired, where $\mathbf{b}_t = 1$ iff its respective transition variable $\mathbf{v}_t > 0$ (that is, the transition is fired with $\alpha>0$). The objective function can then be modified by multiplying it by a constant $c$ and subtracting each boolean variable. The constant $c$ must be large enough to ensure the maximal solution size ($|T|$) does not interfere with the maximal token mass. In this way, the solution size is prioritized between solutions with the same objective value and larger solutions are penalized.

We add appropriate constraints to ensure the MILP uses the boolean variables as intended. The constraint $\mathbf{b}_t - \mathbf{v}_t < 1$ forces token mass to be fired through $t$ when $\mathbf{b}_t = 1$. Oppositely, the constraint $\mathbf{v}_t - \mathbf{b}_t \times c \leq 0$ (where $c$ is a `sufficiently large constant') forces $t$ to not be fired if $\mathbf{b}_t = 0$.

Perhaps most importantly, the introduction of $\mathbf{b}$ ensures that one can check if a given solution is a member of the firing set, and, if not, exclude that specific set of transitions from the search space. Consider a solution $\mathbf{v}'$ to the MILP and its corresponding boolean variables, $\mathbf{b}'$, and assume $\mathbf{b}'$ is undesirable as it is not a member of the firing set; we therefore wish to remove solutions with the same support from the search space. This is achievable by adding the following constraint.

\begin{equation*} \label{eq:fs-forbid} 
    2|(\mathbf{b}\mathbf{b}')^+| < |\mathbf{b}^+| + |{\mathbf{b}'}^+| \tag{\ding{73}}
\end{equation*}

The complete maximization algorithm solves the MILP as described above and, if a solution exists, checks whether the solution is in the firing set using the \textsc{Fireable} algorithm (Alg.~\ref{alg:fireable}). If the solution is in the firing set, it is returned. Otherwise, the support of the solution is excluded from the solution space by adding the constraint (\ref{eq:fs-forbid}) and the MILP is run again.

\begin{algorithm}
    \small
    \caption{MILP maximization}\label{alg:max}
    \hspace*{\algorithmicindent} \texttt{MILPMax}($\langle\mathcal{N}, \mathbf{m}_0\rangle$):\\
    \hspace*{\algorithmicindent} \textbf{Input}: a CPN system $\langle\mathcal{N}, \mathbf{m}_0\rangle, p_g$, where $p_g$ is the goal place \\
    \hspace*{\algorithmicindent} \textbf{Data}: a `suitably large' constant $c$ \\
    \hspace*{\algorithmicindent} \textbf{Output}: the Parikh image of the solution firing sequence
    \begin{algorithmic}[1]
        \STATE{ $ex \gets \emptyset$}
        \REPEAT
            \STATE{\textbf{solve} $\exists?\mathbf{v},\mathbf{b}$, which maximizes objective: $c ({}^\bullet \mathbf{v}[p_g] - \mathbf{v}[p_g]^\bullet) - |\mathbf{b}^+|$ and:
                \begin{itemize}
                    \item $C\mathbf{v} \geq \mathbf{0}-\mathbf{m}_0$
                    \item $\mathbf{v} - \mathbf{b} < \mathbf{1}$
                    \item $\mathbf{v} -\mathbf{b} \cdot c \leq \mathbf{0}$
                    \item $2|(\mathbf{b}\mathbf{b}')^+| < |\mathbf{b}^+| + |{\mathbf{b}'}^+|$ for each $\mathbf{b}' \in ex$
                \end{itemize}}
            \IF{$\exists \mathbf{v} \land \texttt{Fireable}(\langle\mathcal{N}, \mathbf{m}_0\rangle, \mathbf{v}^+$)}
                \RETURN{ $\mathbf{v}$}
            \ELSIF{$\exists \mathbf{v} \land $ \textbf{not} \texttt{Fireable}$(\langle\mathcal{N}, \mathbf{m}_0\rangle, \mathbf{v}^+$)}
                \STATE{$ex \gets \{\mathbf{b}\} \cup ex$}
            \ENDIF
        \UNTIL{$\not\exists \mathbf{v}$}
        \RETURN{\textbf{false}}
    \end{algorithmic}
\end{algorithm}

It is also possible to use (\ref{eq:fs-forbid}) to request the first $n$ solutions which use different transition sets, $n \in \mathbb{Z}^+$ by excluding previously found solutions in addition to solutions which are not in the firing set.
\section{Results.} \label{sec:results}

Our findings are split into two parts: first, we analyze the maximum carbon efficiency of a well-studied metabolic pathway (the pentose phosphate pathway \cite{patra2014pentose}) at various levels of complexity; second, we compare the average running times between the polynomial algorithm and the MILP on both synthetic and chemical data.

The code for \textsc{AtLeastReachable} with \textsc{BinarySearch}, \textsc{MILPMax}, and benchmarking is provided on Github: \url{https://github.com/a2390yu/cpns-a}, alongside the example files used to construct and analyze the pentose phosphate pathway.

\subsection{Pentose phosphate pathway.}

We selected the pentose phosphate pathway (PPP), in particular the non-oxidative segment~\cite{lehninger2004lehninger}, as the target of maximal yield analysis.
The PPP is well suited for our purposes as it has clearly defined source and target molecules, converting ribulose-5-phosphate (R5p) into fructose-6-phosphate (F6p) in the presence of water.
As the PPP pathway is well studied as part of the central carbon metabolism, the yield of F6p is already known.
To make our analysis interesting, we thus consider possible shortcuts, or parallel pathways, induced by the natural promiscuity of the enzymes \cite{tawfik2010enzyme} involved, i.e.~the ability of an enzyme to execute the same reaction on different educt molecules that have similar structure and physicochemical properties.

To obtain chemical networks (and thus Petri nets) which include such promiscuous reactions, we turn to generative models of graph transformation~\cite{andersen2025pathway, andersen2019chemical}.
This approach uses the natural representation of molecules as labeled undirected graphs, and encodes reactions as graph transformation rules.
Crucially, a rule does not require whole molecule(s) as input, but rather can match any partially specified molecule(s); thus a single rule can represent the same reaction executed on different molecules, exactly capturing enzyme promiscuity.

\begin{figure}[ht]
    \centering
    \includegraphics[width=\columnwidth]{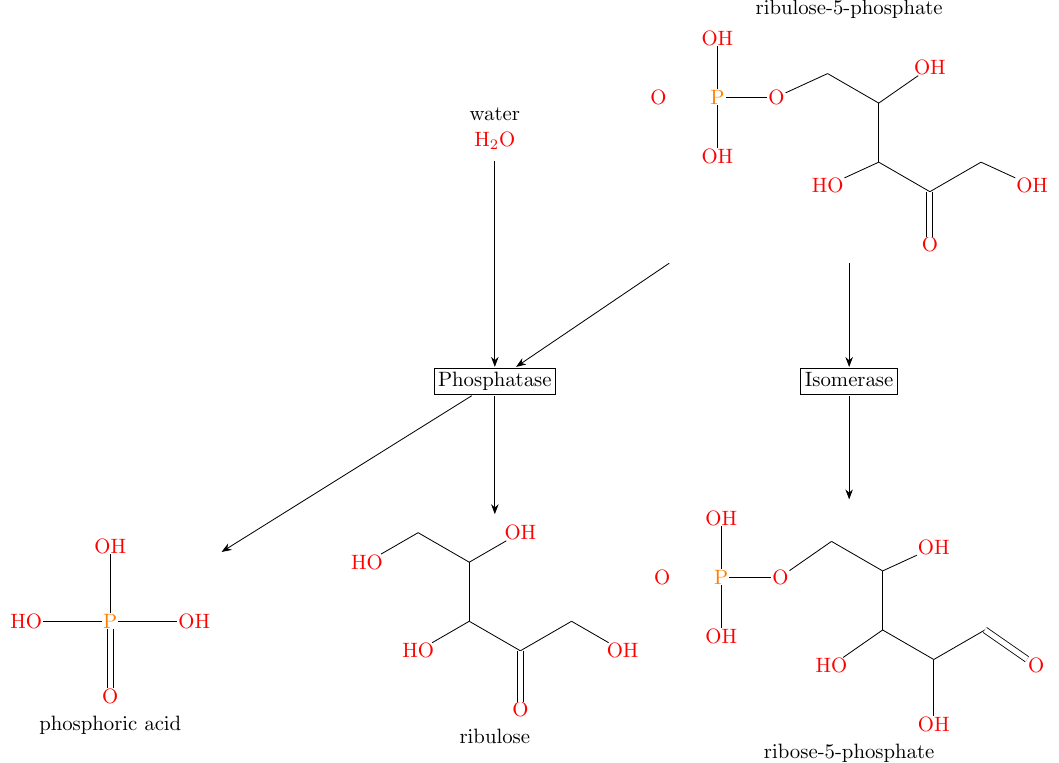}
    \caption{The first expansion step of the PPP, $U_1$. The original universe ($U_0$) consists of only water and R5p (ribulose-5-phosphate). Two graph transformation rules, which model phosphatase and isomerase respectively, are applied; the phosphatase rule applies to R5p and water, producing phosphoric acid and ribulose, while the isomerase rule applies to R5p alone, converting it into ribose-5-phosphate. Thus $P_1$ is composed of three products. The target F6p has not yet been produced.}\label{fig:ppp-1}
\end{figure}

Constructing chemical reaction networks by graph transformation models gives us control over the size, and consequently, the complexity of the network.
In our case, we consider a simple step-wise expansion.
Initially only the source molecules of PPP (R5p and water) are considered in the universe $U_0$.
Next, all possible graph transformation rules are applied, and all product graphs (molecules) $P_1$ are included in the universe at step $1$, $U_1=U_0\cup P_1$.
The illustration of the first expansion step (i.e., the application of all transformation to all molecules in $U_0$) is given in Fig.~\ref{fig:ppp-1}.

Further expansions follow the same procedure, e.g.~$U_2=U_1\cup P_2$, etc.
The target compound, F6p, first appears in $U_3$.

As mass preservation is a core characteristic of chemical systems, the maximum yield of a molecule is naturally upper bounded.
In the case of the PPP, the bound is given by the number of available carbon atoms, all of which come from the source R5p.
The maximum yield of F6p, $100\%$ \emph{carbon efficiency}, is thus achieved when all carbon atoms of R5p end up in F6p molecules.
R5p contains $5$ carbon atoms, meanwhile the target F6p contains $6$.
A token mass of $1$ on R5p can therefore become at best $\frac{5}{6}=0.8\overline{3}$ token mass on F6p, corresponding to $100\%$ carbon efficiency.

Since F6p $\not\in U_0, U_1, U_2$, we examine firstly $U_3$, the third expansion. Starting with the marked CPN $(U_3, \mathbf{m}_0=(H2O: 1, R5p: 1))$, we find the maximum token mass attainable on F6p to be 0.5. Here, $3$ carbons from R5p end up in F6p, resulting in a mass of $\frac{3}{6}=0.5$ on F6p. The carbon efficiency, however, uses $3$ of the initial $5$ carbon atoms in R5p, and thus is $\frac{3}{5}=0.6$.

\begin{figure}[h!t]
    \small
    \centering
    \includegraphics[scale=0.5]{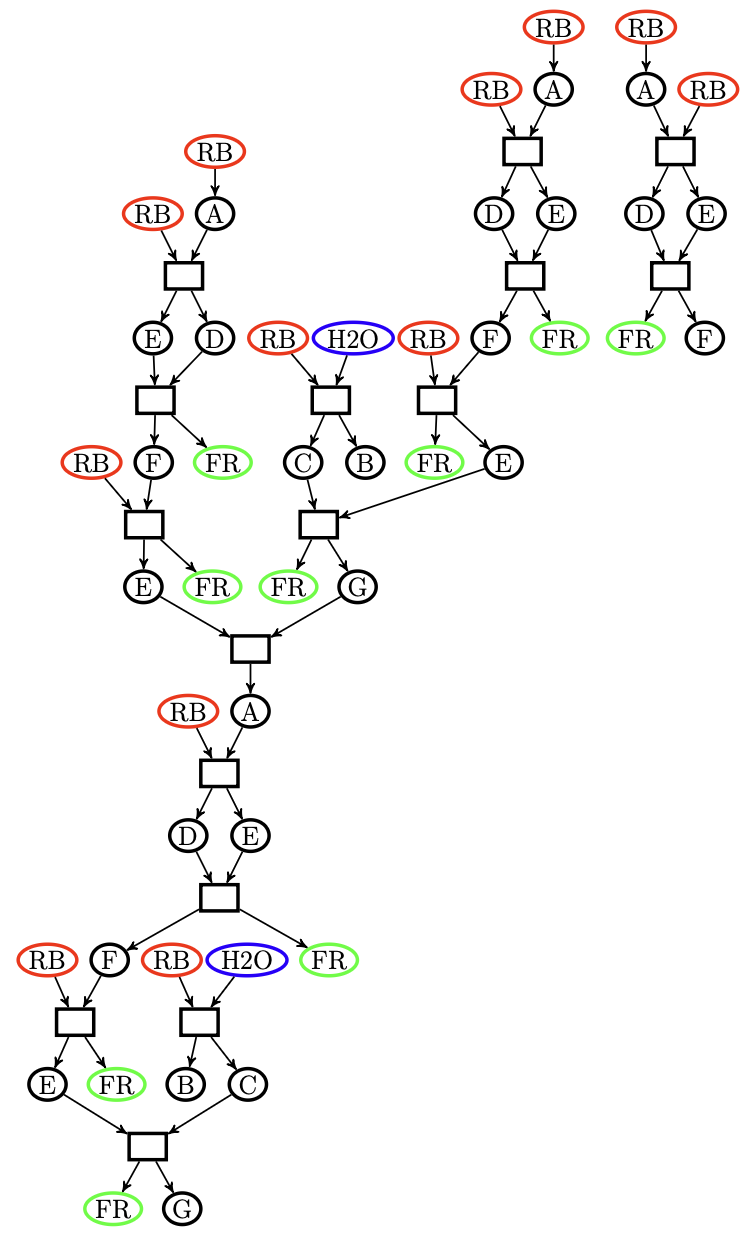}
    \caption{A certificate to the solution given by the MILP which yields  $9$ molecules of F6p (green FR nodes) from $12$ molecules of R5p (red RB nodes) and $2$ molecules of water (blue H20 nodes) in $U_4$ space. See Tab.~\ref{Tab:smiles} for chemical formulae (in SMILES) of all other lettered nodes.}\label{fig:witness-1}
\end{figure}

The simplest PPP pathway hereto-known to achieve maximum carbon efficiency has been studied using discrete models and requires, at its maximum depth, a sequence of five enzymatic reactions \cite{lehninger2004lehninger}, meaning the solution can be found in our $U_5$ (or larger) space. Interestingly, we identified another maximum efficiency pathway using only $U_4$ space, owing to the allowance of limit-reachable pathways. A limit-reachable solution means that optimality is achieved as part of an ongoing process; with a steady supply of R5p, the optimal amount of F6p is continually produced in perpetuum. Since optimality is achieved in the limit, a discrete witness cannot be produced. However, a witness which produces one less molecule of F6p than the theoretical yield is always discretely attainable. The witness to our solution (Fig.~\ref{fig:witness-1}) therefore shows that starting with 12 red RB nodes (R5p) and 2 blue H2O nodes, it is possible to obtain 9 of the expected 10 green FR nodes (F6p), where the final molecule of F6p is only attainable in the limit.



\begin{table*}[t]                   
  \caption{node SMILES strings with ChEBI identifiers as used in Fig.~\ref{fig:witness-1}}
  \setlength{\tabcolsep}{6pt}      
  \begin{tabularx}{\textwidth}{@{}c p{4.3cm} X c@{}}   
    \toprule
    \textbf{Node} & \textbf{Name} & \textbf{SMILES} & \textbf{ChEBI} \\
    \midrule
    RB & ribulose 5-phosphate and xylulose 5-phosphate & \smiles{C(C(C(C(=O)CO)O)O)OP(=O)(O)O} & 17363 \\
    FR & fructose 6-phosphate & \smiles{C(C(C(C(C(=O)CO)O)O)O)OP(=O)(O)O} & 15946 \\
    A & ribose 5‑phosphate & \smiles{C(C(C(C(C=O)O)O)O)OP(=O)(O)O} & 17797 \\
    B & phosphate            & \smiles{O=P(O)(O)O}                    & 18367 \\
    C & ribulose             & \smiles{C(CO)(C(C(CO)O)O)=O}           & 28721 \\
    D & sedoheptulose 7‑P    & \smiles{C(CO)(C(C(C(C(COP(O)(O)=O)O)O)O)O)=O} & 15721 \\
    E & glyceraldehyde 3‑P   & \smiles{C(C(COP(O)(O)=O)O)=O}          & 29052 \\
    F & erythrose‑4‑P        & \smiles{C(C(C(COP(O)(O)=O)O)O)=O}      & 48153 \\
    G & glycolaldehyde       & \smiles{C(CO)=O}                       & 17071 \\
    \bottomrule
  \end{tabularx}
  \label{Tab:smiles}
\end{table*}

\subsection{Running times.}

\begin{table}[h!]
    \small
    \caption{Benchmarking performed by constructing lattices of various heights and widths; each node is connected to its east and south neighbours (if they exist). A random 10\% of resources are given 1.0 token mass; a random node is selected as the goal. Average running time in seconds is taken over 100 repetitions of each algorithm with aforementioned randomized starting conditions. \textsc{BinarySearch} precise to three decimals. ALR with BS is the average time it takes \textsc{AtLeastReachable} with \textsc{BinarySearch} to determine maximum token mass. ALR per iter.~is the average time taken per call to \textsc{AtLeastReachable} alone. Our implementation of \textsc{MILPMax} automatically terminates after excluding 400 transition sets from the solution space (in the interest of time).}
    \begin{tabularx}{\columnwidth}{XXXXX}
        \toprule
            \textbf{Lattice dimensions}
            & \textbf{ALR with BS}
            & \textbf{ALR per iter.}
            & \textbf{MILP} \\
        \midrule
        5x5 & 0.56748 & 0.04974 &  0.00238\\
        5x6 & 1.07829 & 0.08911 & 0.00333\\
        5x7 & 1.83007 & 0.14571 & 0.00477\\
        5x8 & 2.76771 & 0.22071 & 0.00575\\
        5x9 & 3.97575 & 0.31629 & 0.00648\\
        5x10 & 5.57610 & 0.43906 & 0.00814\\
        6x6 & 1.93298 & 0.15415 & 0.00471\\
        6x7 & 3.17880 & 0.25209 & 0.00648 \\
        6x8 & 4.86553 & 0.40546 & 0.00822\\
        6x9 & 7.17460 & 0.59788 & 0.00964\\
        6x10 & 10.12921 & 0.84410 & 0.01226\\
        10x10 & 196.01608 & 14.15279 & 0.13745\\
        20x20 & 17229.93054 & 1077.54412 & 4.18984\\
        \bottomrule
    \end{tabularx}
    \label{Tab:benchmarks}
\end{table}

\begin{table}[h!]
    \small
    \caption{Small network made up of 46 places and 50 transitions based on the metabolism of E. coli. A percentage of all places is randomly selected and given an initial 1.0 token mass. A single goal place is randomly selected. The times are taken in the same way as Tab.~\ref{Tab:benchmarks}. Both algorithms perform worse when less resources are available initially (the smaller the firing sets are), but this effect is far more pronounced for \textsc{MILPMax}.}
    \begin{tabularx}{\columnwidth}{XXXX}
        \toprule
            \textbf{\% of resources}
            & \textbf{ALR with BS}
            & \textbf{ALR per iter.}
            & \textbf{MILP} \\
        \midrule
        50\% & 3.50518 & 0.21907 & 4.08544 \\
        60\% & 3.58185 & 0.22386 & 1.47033 \\
        75\% & 3.41573 & 0.21348 & 0.64617 \\
        80\% & 3.07217 & 0.19201 & 0.33828 \\
        85\% & 2.95147 & 0.18446 & 0.02551 \\
        90\% & 2.95707 & 0.18481 & 0.00908 \\
        \bottomrule
    \end{tabularx}
    \label{Tab:benchmarks-avg}
\end{table}

Although the decision algorithm \textsc{AtLeastReachable} runs in polynomial time, the overhead created by constructing and solving multiple LPs means that it often runs slower than the MILP algorithm. More precisely, the for-loop (line \ref{alr:line:for-loop}) of \textsc{AtLeastReachable} runs for each transition in $T'$, which is initially set to $T$ and thus runs in $O(T)$ time. The while-loop (line \ref{alr:line:while-loop}) which encases the for-loop also operates in $O(T)$ time (when lines \ref{alr:line:sol+}--\ref{alr:line:fslim} only decrement $|T'|$ by 1). Thus, both loops combined result in $O(T^2)$ calls to the LP. In order to speed up the running time, one could parallelize the LP construction within the for-loop.

Additionally, since \textsc{AtLeastReachable} only decides reachability, the algorithm \textsc{AtLeastReachable} must be run multiple times itself in order to target the maximum achievable token mass on the goal place. In practice, using \textsc{AtLeastReachable} with \textsc{BinarySearch} can run significantly slower than \textsc{MILPMax}, as can be seen by the recorded times in Table \ref{Tab:benchmarks-avg}.

However, there are cases when \textsc{AtLeastReachable} with \textsc{BinarySearch} runs faster than \textsc{MILPMax}. One example occurs when no realizable solution exists, but many non-realizable solutions exist. In such a scenario, \textsc{AtLeastReachable} with \textsc{BinarySearch} has an advantage, as \textsc{AtLeastReachable} is called the same number of times whether a realizable solution exists or not. On the other hand, \textsc{MILPMax} will continue to iterate through, in the worst case, the entire powerset of $T$.

Another example similarly occurs when there exist many non-realizable solutions with larger objective functions than any realizable solution. In that scenario, \textsc{MILPMax} must again iterate through all such non-realizable solutions with higher objective function values before it can reach the realizable solution(s).

\textsc{MILPMax} tends to perform better on CPNs with large initial markings (i.e.~many resources available) or many realizable paths to the goal place, whereas \textsc{AtLeastReachable} performs better when the initial marking is scarce (i.e.~resource scarcity) and there are few realizable paths which lead to the goal place (see Table \ref{Tab:benchmarks-avg}).

\section{Summary.} \label{sec:summary}

Continuous Petri nets (CPNs) have established applications in biological modeling. We present a polynomial time algorithm (\textsc{AtLeastReachable}) for deciding if a minimum amount of token mass can be put on a single goal place from an initial marking. By using \textsc{AtLeastReachable} with \textsc{BinarySearch}, one can pinpoint the maximum amount of token mass achievable on the goal place in logarithmic polynomial time.

However, due to the fact that \textsc{AtLeastReachable} constructs and runs multiple LPs and then sums each solution, the final aggregate solution is large and contains spurious transitions. As a result, for non-trivial CPNs, it can be very difficult to identify which transitions actively contribute to the goal as they may be buried among transitions which are fired meaninglessly.

We therefore present a second algorithm which uses MILP to maximize yield while prioritizing solutions of minimum size and ensuring causal soundness; because MILP is NP-complete, the theoretical polynomial runtime of \textsc{AtLeastReachable} is lost. However, in practice, we find that the MILP solution often runs much faster than \textsc{AtLeastReachable}, which we believe is due in part to the overhead of having to construct and solve multiple LPs in order to construct the final solution.

Lastly, we provide an application case study which uses CPNs to analyze the carbon conversion efficiency of the pentose phosphate pathway. We identify new, limit-reachable, solutions with optimal yield of the target molecule, which could not have been discovered using classical discrete methods.
The limit-reachable results are highly relevant, as they essentially capture the ability to maintain an optimal yield of the target molecule under steady supply of the source molecule, a natural condition in a metabolic setting.






\bibliographystyle{siamplain}
\bibliography{bibliography}

\end{document}